\documentclass[aps,prb,preprint,groupedaddress,citeautoscript,nopacs]{revtex4}

\usepackage{graphicx}
\usepackage{color}
\usepackage{amsfonts}
\setcitestyle{super}

\begin{document}

\title{Emergent topological phenomena in thin films of pyrochlore iridates}

\author{Bohm-Jung Yang$^{1}$ and Naoto Nagaosa$^{1,2}$}

\affiliation{$^1$ RIKEN Center for Emergent Matter Science (CEMS), Wako, Saitama 351-0198, Japan}

\affiliation{$^2$ Department of Applied Physics, University of Tokyo, Tokyo 113-8656, Japan}

\date{\today}

\begin{abstract}
{\bf Two quintessential ingredients governing the topological invariant of a system
are the dimensionality and the symmetry of the system. Due to the recent development
of thin film and artificial superstructure growth technique,
it is possible to control the dimensionality of the system, smoothly between
the two-dimensions (2D) and three-dimensions (3D).
In this work we unveil the dimensional crossover of
emergent topological phenomena in correlated topological materials.
In particular, by focusing on the thin film
of pyrochlore iridate antiferromagnets grown along the [111] direction, we demonstrate that
it can show giant anomalous Hall conductance, which is proportional to
the thickness of the film,
even though there is no Hall effect in 3D bulk material.
In addition, we uncover the emergence of a new topological phase,
whose nontrivial topological properties are hidden in the bulk insulator but manifest only in thin films.
This shows that the thin film of topological materials is a new
platform to search for unexplored novel topological phenomena.}
\end{abstract}

\maketitle
Dimensionality and symmetry of a system
govern the topological nature of a band insulator.~\cite{Ryu1, Ryu2, Kitaev, TI_Qi1}
For instance, in two dimensional (2D) electronic systems
with broken time reversal symmetry (TRS),
band insulators can be classified by the quantized Hall conductance,
which generally has the form of
$\sigma_{xy}=\frac{e^{2}}{h}N$ where
$e$ is the elementary charge and $h$ is the Planck constant.~\cite{Haldane}
Since $N$ can be any integer number, there are infinite number of ways
distinguishing band insulators.
However, when 2D and three dimensional (3D) electronic systems obey TRS,
there are only two different ways to discern band insulators based
on the $Z_{2}$ topological index.~\cite{TI_Kane1, TI_Kane2,FuKaneMele, MooreBalents, Roy, Essin}
On the other hand, if 3D electronic systems break
TRS, there is no way to differentiate band insulators
since one insulator can always be deformed to the other insulator adiabatically.
Considering such an intimate relationship between the topological property
of band insulators and the dimensionality (and also the symmetry),
one natural question is
how the dimensional crossover from 2D to 3D limit occurs in a thin film structure.
In fact, it is recently proposed that in systems with TRS,
if we make a thin film by stacking 2D topological insulators, the topological nature
of the film dramatically changes depending on the parity of the number of stacked 2D layers.
Such an oscillating topological property of the film eventually gives rise
to a topologically nontrivial insulator in the 3D bulk limit.~\cite{crossover}

The main purpose of this work is to establish
such a fundamental relationship between the dimensionality and the topological
nature of a system in correlated topological materials.
In particular, we focus on the evolution of topological properties
of a transition metal oxide (TMO) with strong spin-orbit coupling in its thin film form.
The motivation to study TMO with large spin-orbit coupling is two-fold.
At first, in this system, due to the strong spin-orbit coupling and the resultant enhanced electron correlation,
new types of topological phases can appear.
For instance, Weyl semi-metallic states (Weyl-SM) which have several Fermi points with 3D Dirac-like dispersion relation
around them
are proposed in several magnetic systems such as HgCr$_{2}$Se$_{4}$~\cite{Fang}
and pyrochlore iridate compounds R$_{2}$Ir$_{2}$O$_{7}$
(R=rare earth elements)~\cite{Wan}. Due to the nontrivial topological properties of Fermi points,
a Weyl-SM has Fermi-arcs on the surface and can show anomalous Hall effect (AHE).
Moreover, in spinel compounds AOs$_{2}$O$_{4}$ (A=Ca, Sr)~\cite{Wan2} as well as
in pyrochlore iridates R$_{2}$Ir$_{2}$O$_{7}$~\cite{Wan}, axionic insulators
are proposed, which can show unusual magneto-electric effects.~\cite{Turner}
The other reason why TMO films are particularly important is due to
the recent technical development to grow artificial superstructures using TMO.~\cite{TMO1}
Since it is possible
to prepare layered structures of TMO in atomic precision and control
material properties of superstructures,
a superlattice or a thin film of TMO is an ideal playground
to search for novel topological phases in thin films.~\cite{TMO2}

Considering recent advances in the study of Ir-based 5d TMOs,~\cite{BJKim1, BJKim2}
here we focus on the topological properties of
pyrochlore iridate thin films.
Since there are natural cleavage planes along the [111] or
its symmetry equivalent directions,
we investigate the dimensional crossover of [111] films varying
the number of layers. The main results of this study are as follows.
Most importantly, we find that the [111] thin film of pyrochlore iridates
can show giant AHE. In particular, in the case of ultrathin films,
the maximum magnitude of the anomalous Hall conductance is proportional
to the total number of layers . In addition, we demonstrate
the existence of a hidden topological phase between
the Weyl-SM and the fully-gapped antiferromagnetic insulator (AFI) in thick films
close to the bulk limit.
Contrary to the common belief that Weyl-SM directly turns into
a trivial AFI as the strength of the local Coulomb interaction ($U$) increases,
we show that the topological properties of the Weyl-SM phase
transfer to its neighboring fully gapped insulator
for a finite range of $U$ values and bring about a new topological state in thin films.
The nontrivial topological properties of the hidden topological phase
are characterized
by in-gap states localized on the surfaces and giant AHE.
These results are summarized in the phase diagram of Fig. 2b.

\section*{Results}

{\bf 3D bulk properties.}
The pyrochlore lattice
can be viewed as a face-centered cubic lattice with a tetrahedral unit cell as shown in Fig. 1a.
In pyrochlore iridates, each Ir$^{4+}$ ion sits on every corner of a tetrahedron.
The candidate non-interacting ground states proposed up to now include
a 3D $Z_{2}$ topological insulator (TI)~\cite{Balents}
and a semi-metal with quadratic band crossing at the Fermi level (quadratic-SM)~\cite{Yang, William2, EG}
as well as conventional metallic states.~\cite{William2}
On the other hand, once electron-electron interactions are considered, an antiferromagnetic
state with the all-in/all-out (AIAO) type local spin configuration can appear
for $U>U_{c1}$ according to the recent theoretical and experimental studies.~\cite{Wan,William2,William1,Arima,Ara}
In the AIAO state, four spins residing on a tetrahedron
always point to or away from the center of the tetrahedron as shown in Fig. 1b.
The AIAO antiferromagnetic state supports two different types of ground states
depending on $U$, i.e., a Weyl-SM and a AFI as described in Fig. 1c.
In fact, Weyl-SM phase has topological stability because each Weyl point (WP)
carries an integer chiral charge. Therefore only when every WP is pair-annihilated
by colliding with another WP with opposite chiral charge at $U=U_{c2}$,
a gapped insulator can be obtained.
In this way, the phase diagram of the 3D bulk system shows
the continuous evolution from the non-magnetic ground state ($0<U<U_{c1}$)
to the Weyl-SM ($U_{c1}<U<U_{c2}$) and finally to the fully gapped
AFI ($U>U_{c2}$) as shown in Fig. 1c.

To describe the evolution of the ground state of the 3D bulk as well as the thin films consistently,
we numerically study a lattice Hamiltonian
whose detailed structure is shown in Methods section.
To obtain the simple but realistic Hamiltonian capturing the key physical properties
of pyrochlore iridates, it is assumed that the low energy physics of the system can be
described by focusing on the doublet of Ir$^{4+}$ with the effective total angular momentum $J_{\rm{eff}}=1/2$,
which is valid when the spin-orbit coupling is large enough.
The transition from a nonmagnetic state to the antiferromagnetic state can be
described by adding a local Hubbard-type interaction $U$.
According to the recent theoretical studies of the lattice Hamiltonian,~\cite{William2, William1, Ara} it is shown that when the
quadratic-SM is the non-magnetic ground state,
the local Coulomb interaction can give rise to the AIAO type magnetic ground state,
consistent with the first-principle calculations and experiments.
Considering the previous theoretical studies,
we choose the parameters of the lattice Hamiltonian in a way that
the quadratic-SM becomes the non-magnetic ground state of the system.
Moreover,
for the description of the local spin ordering pattern consistent with AIAO state,
we introduce a vector $\vec{m}_{i}$ at each lattice site $i$,
and use the magnitude of the local spin moment $|\vec{m}_{i}|\equiv m$
as a tunable parameter, representing the strength of $U$.
In this approach, the quadratic band touching point of the quadratic-SM
locating at the $\Gamma$ point of the Brillouin zone (BZ)
immediately splits into 8 WPs once $m>0$.
In this Weyl-SM phase, a pair of WPs are aligned along the [111] direction
and the other 3 pairs of WPs are aligned along the other symmetry equivalent directions
as shown in Fig. 1d.
Moreover, the pair of WPs
existing along the [111] direction have opposite chiral charges
and the distance between them increases as $m$ increases.
When $m=m_{c1}$, the pair reach a BZ boundary simultaneously
and a pair-annihilation occurs.
Because of the cubic symmetry, the same thing happens for the other 3 pairs
of WPs leading to AFI.
Therefore $m=0$ ($m=m_{c1}$) is basically equivalent to the critical point at $U=U_{c1}$ ($U=U_{c2}$)
in Fig. 1c.

{\bf Hall conductance of a Weyl-SM phase and cubic symmetry.}
The nontrivial physical response triggered by WPs is the intrinsic AHE.
This is because a WP behaves like a fictitious magnetic monopole,
which generates Berry gauge flux.
Since the anomalous Hall conductivity is given by the integration of Berry gauge flux in the momentum space,
Weyl-SM phase can show AHE.
For example, the Hall conductivity of a Weyl-SM having only a pair of WPs at $\pm\textbf{k}_{0}=(0,0,\pm k_{0})$
is given by $\sigma_{xy}=\frac{e^{2}}{h}\frac{2k_{0}}{2\pi}$,
where $2k_{0}$ is the distance between two WPs.~\cite{Wan, Ran,Burkov}
To understand the relation between the AHE and the distribution of WPs,
it is convenient to introduce a Chern vector $\vec{C}=(C_{x},C_{y},C_{z})$, which is defined
as $\sigma_{ij}=\frac{e^{2}}{2\pi h}\epsilon_{ijk}C_{k}$ where $\epsilon_{ijk}$ is the fully-antisymmetric tensor
with $\epsilon_{123}=1$.~\cite{3DQHI,Ran} Therefore, for a pair of WPs
located at $\pm\textbf{k}_{0}=(0,0,\pm k_{0})$, the corresponding Chern vector would be $\vec{C}=(0,0,2k_{0})$
and the location of the two Weyl points are at $\textbf{k}=\pm\frac{1}{2}\vec{C}$.
As the distance between
the two Weyl points increases, the Hall conductance of the system grows
until the two Weyl points are pair-annihilated at the zone boundary. Once
the pair-annihilation occurs,
the system turns into an insulator with the quantized Hall conductivity
of $\sigma_{xy}=\frac{e^{2}}{ha_{z}}$ ($a_{z}$ is the lattice constant along the $z$-direction),
which is nothing but the 3D Chern insulator (or the 3D quantum Hall insulator).

On the other hand, in the case of the Weyl-SM phase in pyrochlore iridates,
there are 4 pairs of WPs due to the cubic symmetry of the system.
Since a Chern vector $\vec{C}$ can be assigned
to each pair of WPs, there are in total four Chern vectors $\vec{C}_{1,2,3,4}$ in the system.
Note that $\pm\frac{1}{2}\vec{C}_{1,2,3,4}$ correspond to the location of the WPs
in the momentum space.
However, since all WPs are related by the cubic symmetry, $\sum_{i=1}^{4}\vec{C}_{i}=0$.
Therefore the total Hall currents
should be zero in the 3D bulk system.
However, it should be stressed that although the net Chern vector of the system adds up to zero,
the Weyl-SM phase of the pyrochlore iridates is still topologically nontrivial
because there exist finite Chern vectors antiferromagnetically aligned,
which are associated with the WPs.
Once the cubic symmetry is broken,
the incomplete cancelation of Chern vectors can generate nontrivial responses,
which is the fundamental origin of emergent topological phenomena in thin films.

Moreover, we find that, surprisingly, even when all WPs are pair-annihilated and the system becomes an insulator,
nontrivial topological responses still persist near the phase boundary between
the Weyl-SM and the AFI as long as the cubic symmetry is broken.
At the critical point ($m=m_{c1}$) where all WPs hit the BZ boundary,
the magnitude of each Chern vector reaches its maximum value.
In contrast to the case of the system having only one Chern vector
where the Chern vector with the maximum magnitude
mediates the transition from a Weyl-SM to the 3D Chern insulator with the quantized Hall conductance,
the antiferromagnetically aligned Chern vectors in pyrochlore iridates
cannot support quantized topological number in the bulk insulator.
However, once the cubic symmetry of the system is broken, for example, by making
a thin film, the surface states associated with the antiferromagnetically aligned Chern vectors
can induce nontrivial topological properties.
For example, 2D conducting channels can appear
on the surfaces or in the domain walls of the bulk insulator.
Moreover, in the case of thin films having two parallel
surfaces on the top and bottom layers,
the coupling between two surface states can induce giant anomalous Hall conductance.
Namely, the nontrivial topological property originating from the antiferromagnetically aligned Chern vectors,
which is hidden in the 3D bulk insulator,
can be manifested as the emerging topological properties of the thin films.

{\bf Topological properties of [111] thin films.}
An alternative way to view the pyrochlore lattice is that
it is composed of alternating kagome and triangular lattices stacked along the [111]
direction as shown in Fig. 2a. Because of such a peculiar lattice structure, the [111] plane provides a
natural direction for film growth. Considering the tetrahedral unit cell,
a bilayer composed of a pair of neighboring kagome and triangular lattices forms
a basic building block for [111] films.
In the following, we examine how the topological properties of thin films
evolve as the number of bilayers ($N_{b}$) increases
by computing the Hall conductance.
Since the finite thickness of the film induces a small gap at the WPs,
the bulk states do not touch the Fermi level for any $m>0$ in general, hence
the Hall conductance is expected to be quantized. However,
the nontrivial topological property of the bulk states
supports surface states in the gap, which disturb the quantization of the Hall conductance in films.
To understand the intrinsic topological property of thin films caused by the interplay between the bulk and surface states,
we compute the Hall conductance of the system in two different ways.
One way is to follow the conventional definition of the
Hall conductance $G_{xy}$ which is nothing but the integral
of the Berry curvature of the occupied states below the Fermi energy.
Because of the surface states in the gap, $G_{xy}$ is normally not quantized
if the film is not ultrathin ($N_{b}<4$).
In addition to $G_{xy}$, we define the maximum Hall conductance $G^{\text{max}}_{xy}$
by imposing the half-filling condition at each momentum $\textbf{k}$ locally
and adding the contribution of locally half-filled bands
over the entire Brillouin zone.
Since the Fermi level can be in the gap locally in each $\textbf{k}$ in this case,
$G^{\text{max}}_{xy}$ can be quantized.

In Fig. 2b, we show
the evolution of the phase diagram of films as $N_{b}$ varies.
Surprisingly, Fig. 2b has unexpected rich structures.
At first, when $0<m<m_{c1}$ corresponding
to the Weyl-SM phase in the bulk limit, the maximum Hall conductance $G^{\text{max}}_{xy}=L_{z}\sigma_{xy}$
($L_{z}=N_{b}a_{z}$ is the length of the film)
increases monotonously
as $m$ increases, and eventually reaches the quantized value of $G^{\text{max}}_{xy}=\frac{e^{2}}{h}N_{b}$
near $m=m_{c1}$.
This change is quite distinct from the oscillating $G^{\text{max}}_{xy}$ predicted for films under the periodic boundary condition (BC)
as shown in Figs. 3a, b.
To understand such a strong BC dependence of $G^{\text{max}}_{xy}$,
it is noted that the change of $G^{\text{max}}_{xy}$ describes a topological
phase transition induced by an accidental gap-closing. Therefore
BC dependence of $G^{\text{max}}_{xy}$ indicates
the fact that gap-closing condition strongly depends on BC.
In fact, a gapless WP of the bulk Weyl-SM phase turns into
a gapped 2D Dirac point (GDP) in films due to the finite size effect.
By changing $m$, whenever an accidental gap-closing
occurs at a GDP, $G^{\text{max}}_{xy}$ shows a discontinuous change of $\Delta G^{\text{max}}_{xy}=\pm\frac{e^{2}}{h}$.~\cite{Oshikawa}
Under the periodic BC, it can be shown that an accidental gap-closing can happen at each GDP separately
by changing one external parameter that determines the location of bulk WPs. (See Methods.)
Since the sign of $\Delta G^{\text{max}}_{xy}$ varies depending on the location of WPs,
the net contribution from all eight GDPs gives rise to the oscillating $G^{\text{max}}_{xy}$.
On the other hand, under the open BC corresponding to the real situation of thin films,
in general it is impossible to induce a gap-closing at a single GDP separately as shown
in Methods section.
In fact, under the open BC, the variation of an external
parameter merely induces an overall shift of the energy spectrum near each GDP,
hence the gap between the valence and conduction bands is fixed irrespective
of the external parameter.
However, even in this case, if the GDP has double degeneracy,
gap-closing is possible.
This is because when the energy spectrum of each of these two GDPs
is shifted to the opposite direction, band touching can occur between
the conduction band of one GDP and the valence band of the other GDP. (See Methods.)
It is worth to note that in the case of [111] films,
there always are one pair of WPs aligned along the $z$-direction,
which is a unique property of [111] films.
Under the open BC, both of these WPs are projected to the
center of the surface BZ while the other six WPs
are projected to other six inequivalent points in the BZ.
Therefore successive gap-closing is possible only at
the $\Gamma$ point leading to
monotonous stepwise increment of $G^{\text{max}}_{xy}$.
In terms of Chern vectors, this means that among the four Chern vectors $\vec{C}_{1,2,3,4}$,
only $\vec{C}_{1}$ which is parallel to the $z$-direction
can contribute to $G^{\text{max}}_{xy}$.
Since the maximum Hall conductance achievable
through a pair of WPs related with $\vec{C}_{1}$ is equal to $\frac{e^{2}}{h}N_{b}$,
this limits the upper bound for $G^{\text{max}}_{xy}$ of the film.

Contrary to the case of $G^{\text{max}}_{xy}$,
the quantization of the physical Hall conductance $G_{xy}$
is disturbed by the surface states in the bulk gap.
As shown in Fig. 3b, the quantized $G_{xy}$ can be observed
only in ultrathin films with $N_{b}<4$.
However, the overall change of $G_{xy}$ follows
the variation of $G^{\text{max}}_{xy}$ and a giant Hall conductance $G_{xy}$
can be observed near the critical point $m=m_{c1}$.
Moreover, it is worth to note that a finite Hall conductance always appears when $m_{c1}<m<m_{c2}$.
This clearly shows that there are two types of metallic conducting channels that can be observed
even when the bulk material is insulating with $m_{c1}<m<m_{c2}$.
When there is a surface state touching the Fermi level in the gap,
hence $G_{xy}$ is not quantized as in the case of thick films,
the top and bottom surfaces of the film support metallic states.
On the other hand, when there is no state touching the Fermi level,
hence $G_{xy}$ is quantized as in the case of the ultrathin films, metallic conducting channels can appear
at the domain wall due to the quantized Hall conductance.

{\bf The hidden topological property of the bulk insulator and
emergent topological phenomena of the associated [111] thin films.}
The observation of the giant Hall conductance $G^{\text{max}}_{xy}=\frac{e^{2}}{h}N_{b}$
near the critical point at $m=m_{c1}$
naturally leads to the following question:
how does the system release such large Hall currents in the strong coupling limit
where a topologically trivial Mott insulator is expected?
Since the bulk states are fully gapped for $m>m_{c1}$,
the release of Hall currents is possible only if the film supports intrinsic surface states
which connect the bulk valence and conduction bands.
The nontrivial topological property of the insulator,
which is hidden in the 3D bulk system,
manifests through the emergence of surface states carrying large Hall currents in thin films.

To understand the hidden topological property of the bulk insulator, let us first describe
the band structure and surface spectrum of a [111] film with the thickness close to the bulk limit,
which has the kagome lattice on both the top and bottom surfaces.
As shown in Fig. 4,
when $0<m<m_{c1}=0.29$,
there are three pairs of Fermi arcs in the surface BZ
whose end points correspond to the projected wave vectors of the bulk WPs to the surface.
For $0<m<m_{c1}$, the total surface states, including the contribution
from both the top and bottom surfaces, form a closed loop centered at the $\Gamma$ point
and the size of the loop expands as $m$ increases as shown in Figs. 4b, c.
However, at $m=m_{c1}$,
the Fermi surface touches BZ boundary signaling a Lifshitz transition (Fig. 4d).
Finally, when $m>m_{c1}$, the surface spectrum turns into
two different closed loops centered at two corners of the BZ, respectively,
as shown in Fig. 4e.
It is worth to note that the surface spectrum does not immediately disappear even when $m>m_{c1}$
and persists in the gap.
In terms of the evolution of the surface spectrum, the critical point $m=m_{c1}$ marks the point where
the Fermi surface topology changes from the open Fermi surface (Fermi arcs)
to the closed Fermi surfaces.
Since the states localized on the top and bottom surfaces exist
at different momenta, in general, there is no hybridization between these two surface states.

On the other hand, when the unit bilayers are stacked along the $z$-direction,
the top and bottom surfaces have different lattice structure. Let us assume
that the top (bottom) surface has the triangular (kagome) lattice structure.
One interesting property of the top surface terminated by a triangular lattice
is that there are additional localized states which are not of topological origin.
In fact, because of the lattice geometry, many localized states confined around each elementary hexagonal plaquette of the lattice can appear
in the tight-binding Hamiltonian on the pyrochlore lattice.~\cite{GSS}
When some of the hexagonal plaquettes near the surface
are broken due to the lattice termination at the surface, the states, which were localized
within those plaquettes before they are broken, are liberated from them and constitute a new surface state.
(See Supplementary Note 1 and Supplementary Figure S1 and S2.)
Such lattice geometry-induced surface states (geometrical-SS) can induce remarkable physical consequences.
For example, as shown in Fig. 5e, even when $m>m_{c1}$,
the conduction and valence bands are still coupled
since there are two bands touching near the momentum $K_{2}$.
In fact, these two bands are surface states localized on the top and bottom surface layers, respectively.
By comparing Fig. 4e and Fig. 5e, we can easily see that
the surface states smoothly connected to the valence band
are nothing but the surface states of topological origin related with the Weyl-SM phase (topological-SS).
On the other hand, the other surface state connected to the conduction band
is a geometrical-SS resulting from lattice geometry.
Since these two surface states have different origins, they can
appear at the same momentum, hence can be hybridized.
Such a crossing between two surface states induces a huge change in
$G_{xy}$ and $G^{\text{max}}_{xy}$, finally leading to the trivial AFI at $m=m_{c2}$
beyond which two surface states decouple.
Since the giant Hall currents are released through the crossing of two surface states,
the crossing points of two surface states should accompany pronounced Berry curvature distribution around them.
As shown in Fig. 6 and Fig. 7, such a strong enhancement of the Berry curvature
can be clearly observed near the region
where two surface states are crossing.

Table 1 summarizes the topological properties of [111] thin films
for various surface termination when the bulk
is a fully gapped insulator with $m_{1}<m<m_{2}$.
The point is that the surface terminated by a triangular
lattice has the topological-SS related with the Weyl-SM
and the geometrical-SS related with the lattice structure at
the same time.
On the other hand, the surface terminated by a kagome lattice
only has topological-SS. Notice that when the two surfaces
have different lattice structures, the thin film can
show the large anomalous Hall effect (AHE) when $m_{1}<m<m_{2}$.
As long as at least one surface of the film is terminated by the kagome lattice,
2D surface metallic states appear in the bulk gap.
However, when the film has the triangular lattice on both surfaces,
the topological-SS derived from Weyl-SM are destroyed by the geometrical-SS, hence
the film becomes topologically trivial.

\section*{Discussion}

As described above, the thin films of pyrochlore iridates
offer a unique opportunity in many aspects.
(i) This is the first realistic antiferromagnetic system showing
the quantized AHE due to the scalar spin chirality
{\it without} the uniform magnetization.~\cite{Kagome_QH}
Namely, although the net magnetization vanishes due to the
cancelation of four spin moments on each tetrahedron, the Chern numbers
of Bloch states in momentum space remain nonzero and result
in the quantized Hall conductance.
The antiferromagnetic phase has two degenerate ground states, i.e., AIAO state and its time-reversed partner.
Since the Hall conductances of these two phases have
the opposite sign, metallic conducting channels can appear at domain walls
as long as the domain wall plane is not perpendicular to the surface normal direction of the film.
(ii) The bulk-surface correspondence
in the thin films derived from the hidden topological phase belongs to a new class, i.e.,
the surface states carry large Chern numbers, which in principle can be proportional to
the thickness $L_{z}=N_{b}a_{z}$ of the film, while the 3D bulk system ($L_z \to \infty$)
is not topological and has a zero Chern number.
This apparently contradicting behavior is explained by the fact that
exponentially small overlap, i.e., $\sim \exp(-L_z/\xi)$
($\xi \sim ta/E_G$ is the correlation length determined by
the transfer integral $t$, the band gap $E_G$ and the lattice constant $a$),
between the surface states on the top and bottom produces the Chern numbers,
and hence carries the information about the bulk. In the limit of $L_z \to \infty$,
this overlap can be neglected and the Chern number vanishes.

Now we remark on the relevant real materials.
There are several antiferromagnets with AIAO spin structures
such as Cd$_2$Os$_2$O$_7$ and R$_2$Ir$_2$O$_7$ (R=Nd, Sm, Eu).
Furthermore, in these materials, one can control the strength of electron correlations.
For example, the recent transport and optical studies on
Nd$_{2}$(Ir$_{1-x}$Rh$_x$)$_2$O$_7$ have shown that
insulator-metal transition can be achieved by replacing Ir by Rh.~\cite{Ueda}
Moreover, in this system, large additional
conductivity is observed due to possible conducting channels at domain walls,
not only when the bulk is expected to be a Weyl-SM state but also when it is a gapped insulator.~\cite{Tokura}
Our study clearly shows that the surface states associated with the anti-ferromagnetically aligned Chern vectors
can appear when the bulk system is a gapped insulator as well as when it is a Weyl-SM, consistent with
the experimental observation.
Although the domain wall is different from the open surface in a strict sense,
the emerging metallic states share the same topological origin.
Such a metallic conduction at domain walls
is also predicted in a recent theoretical study.~\cite{Imada}

Finally, let us discuss about the influence of surface disorder in real materials.
At first, when $0<m<m_{c1}$, the complexity of the surface is not crucial
because the only requirement to observe the nontrivial change
of Hall conductance induced by the topological charge of WP is to impose a sharp (hard wall)
boundary condition, which is obviously achievable in experiments.
However, disorder effect can be important when $m>m_{c1}$,
especially related with the Hall conductance of the films.
In contrast to the case of the films associated with the Weyl-SM phase where the change of Hall conductance is induced
by bulk states, in the case of the films derived from the bulk insulator with $m_{c1}<m<m_{c2}$, giant Hall currents
are carried by two surface states whose wave function overlap is exponentially small in thick films.
Therefore once the complexity of the surface is considered,
the Hall currents of the films
can be destroyed, especially, in the case of thick films.
However, the quantized Hall conductance of ultrathin films having large hybridization gap between two surface states
and the giant Hall response near the critical point $m_{c1}$ can remain robustly even in the presence of disorder.
Therefore although the interval between $m_{c1}$ and $m_{c2}$ can be reduced
due to the disorder effect,
the overall structure
of the phase diagram should remain intact against disorder.

To conclude, we have theoretically studied the topological dimensional
crossover phenomena in [111] thin films of pyrochlore iridates.
We found the large anomalous Hall conductance
even in the films associated with a fully gapped insulator in the bulk limit,
which we call the hidden topological phase.
The large Hall currents in the films derived from the bulk insulator phase is due to two surface states, i.e.,
one from the topological reason associated with WPs and
the other from the geometrical reason of the lattice structure.
Our theoretical study is based on the general tight-binding Hamiltonian,
which captures the key intrinsic properties of pyrochlore iridates,
hence can reproduce the essential physical properties
of the 3D bulk system consistent with the first-principle calculations and
experiments.
Considering the growing experimental efforts to make 5d TMO thin films,
it is expected that interesting experimental data on the pyrochlore iridate thin films would be available
in near future, which would also provide useful guidance for more
sophisticated numerical studies such as the first-principle calculations.
This work will pave a way to explore the new emergent topological phenomena
by artificial superstructures such as thin films in both theory and experiment.

\section*{Methods}
{\bf Lattice Hamiltonian for pyrochlore iridates.}
To describe the evolution of the ground state for thin films as well as
the 3D bulk system,
we use the following lattice Hamiltonian proposed in Ref.~\onlinecite{William2};
\begin{eqnarray}
H_{0}=\sum_{\langle i,j\rangle}c^{\dag}_{i}(t_{1}+it_{2}\vec{d}_{ij}\cdot\vec{\sigma})c_{j}
+\sum_{\langle\langle i,j\rangle\rangle}c^{\dag}_{i}(t'_{1}+i[t'_{2}\vec{R}_{ij}+t'_{3}\vec{D}_{ij}]\cdot\vec{\sigma})c_{j},
\end{eqnarray}
where $\sigma_{1,2,3}$ are Pauli matrices describing the effective spin degrees of freedom.
Here it is assumed that the low energy physics of the system can be
described by focusing on the doublet of Ir$^{4+}$ with the effective total angular momentum $J_{\rm{eff}}=1/2$,
which is valid when the spin-orbit coupling is large enough.
$t_{1,2}$ ($t'_{1,2,3}$) indicates the hopping amplitude between
the nearest-neighbor (next nearest neighbor) sites.
The real vectors $\vec{d}_{ij}$, $\vec{R}_{ij}$, and $\vec{D}_{ij}$
describe $\vec{\sigma}$ dependent hopping terms.~\cite{William2}
The transition from a nonmagnetic state to the antiferromagnetic state can be
described by adding $H_{\text{m}}=\sum_{i}c^{\dag}_{i}(\vec{m}_{i}\cdot\vec{\sigma})c_{i}$,
which can be considered as the result of a mean field approximation for the Hamiltonian
with the Hubbard-type local interaction $U$.~\cite{William1}
Here the vectors $\vec{m}_{i}$ describe the local spin ordering patterns compatible with
AIAO state.

{\bf Oscillating Hall conductance of [111] films in the Weyl-SM phase under the periodic BC.}
In Fig. 3a of the main text, we have shown the change of the Hall conductance
$G_{xy}=L_{z}\sigma_{xy}$ (or Chern number)
as $m$ increases under the periodic BC along the $z$-direction,
the surface normal direction of the film.
When $m$ is within the range where the 3D bulk system is in a Weyl-SM phase,
thin films show oscillating Hall conductance.
Here we explain the origin of
the oscillating $G_{xy}$ in terms of Chern vectors.
By definition, only the $z$-component of each Chern vector $C_{i}$ can contribute
to $G_{xy}$.
In a [111] thin film, one of $\vec{C}_{1,2,3,4}$ is parallel to the $z$-direction.
Suppose that $\vec{C}_{1}$ is parallel to the $z$-direction.
Then the z-components of other Chern vectors $\vec{C}_{i\neq1}$ is equal to $-\frac{1}{3}C_{1,z}$.
Since the sign of $C_{i\neq1,z}$ is opposite to that of $C_{1,z}$,
their contributions to $G_{xy}$ also have the opposite sign.
Let us first consider the contribution of $\vec{C}_{1}$ to $G_{xy}$.
Because of the finite-size effect, the momentum $k_{z}$ is discretized and
only the discrete momenta within
the range of $[-\frac{1}{2}C_{1,z},\frac{1}{2}C_{1,z}]$ can contribute to $G_{xy}$.
On the other hand, in the case of $\vec{C}_{i\neq1}$, the discrete momenta
within the range of $[-\frac{1}{6}C_{1,z},\frac{1}{6}C_{1,z}]$ can contribute
to $G_{xy}$.
Since the number of discrete momenta within
the range of $[-\frac{1}{2}C_{1,z},\frac{1}{2}C_{1,z}]$
is not always equal to the triple of the number of discrete momenta within
the range of $[-\frac{1}{6}C_{1,z},\frac{1}{6}C_{1,z}]$,
the oscillating $G_{xy}$ appears.
Notice that the maximum magnitude of the 3D conductivity $\sigma_{xy}=\frac{2e^{2}}{ha_{z}}\frac{1}{N}$
vanishes in the bulk limit ($N\rightarrow\infty$), consistent with the constraint of $\sum_{i=1}^{4}\vec{C}_{i}=0$
in the bulk system. Such an oscillating $\sigma_{xy}$ is generally expected
in the Weyl-SM phase with a zero total Chern vector,
when the system has a thin film structure
under the periodic boundary condition.

{\bf Accidental gap closing in thin films of the Weyl-SM phase.}
One peculiar property of the pyrochlore iridate [111] film is that the accidental
gap-closing condition strongly depends on the boundary conditions. Namely,
in the case of the periodic boundary condition,
an accidental band crossing at a single Weyl point can be induced by changing the magnitude of the local spin moment ($m$).
On the other hand, under the open boundary condition, a gap-closing at a single Weyl point cannot be induced
by controlling $m$. Interestingly, in the case of the film with open boundary condition,
only when a pair of Weyl points are aligned along the surface normal direction, band crossing occurs.
Such a strong boundary condition dependence of the gap-closing condition gives rise
to stark difference in the change of the Hall conductance of a film depending
on the boundary conditions.

The origin of the boundary condition dependence of the gap-closing condition is
as follows. At first, under the periodic boundary conditions,
the typical eigenfunction and the corresponding eigenvalues of a Weyl fermion can be written as
\begin{displaymath}
\psi(\vec{r})=Ce^{iqz}e^{i\vec{k}_{\perp}\cdot\vec{r}_{\perp}}
\left(\begin{array}{c}
k_{x}-ik_{y}\\
\varepsilon/\upsilon-q
\end{array}\right),
\quad \varepsilon=\pm\upsilon\sqrt{k^{2}_{x}+k^{2}_{y}+q^{2}}
\end{displaymath}
where $q$ is the momentum component normal to the surface,
$\vec{r}_{\perp}=(x,y)$, $\varepsilon$ is the energy eigenvalue and $\upsilon$
is the velocity of the Weyl fermion. To determine the
discrete value of the momentum $q$, we consider the following periodic boundary condition,
$\psi(z+L_{z}/2)=e^{i\phi}\psi(z-L_{z}/2)$. It is worth to note that in general two wave functions at
$z=-L_{z}/2$ and $z=L_{z}/2$ can be different up to the overall U(1) phase $\phi$.
This is because when we derive the low energy Hamiltonian near the Weyl points,
we assume $\Phi(\textbf{r})\sim e^{i\textbf{k}_{W}\cdot \textbf{r}}\psi(\textbf{r})$
where $\textbf{k}_{W}$ is the location of the Weyl point. Here $\Phi$
is the full lattice eigenstate and $\psi$ is the low energy state near the Weyl point.
Therefore if we impose $\Phi(z+L_{z}/2)=\Phi(z-L_{z}/2)$, it is equivalent to
$\psi(z-L_{z}/2)=\psi(z+L_{z}/2)e^{i k_{W,z}L_{z}}$.
Since the location of the Weyl point $\textbf{k}_{W}$ depends on the local spin moment $m$, we can obtain
$\psi(z+L_{z}/2)=e^{i\phi}\psi(z-L_{z}/2)$ with $\phi=\phi(m)$.
This condition immediately gives rise to discrete $q=\frac{2n\pi+\phi}{L_{z}}$ value with $n$ integer,
and the corresponding energy eigenvalues
given by
\begin{eqnarray}
\varepsilon_{+}(q,\vec{k}_{\perp}=0)/\upsilon=q=\frac{2n\pi}{L_{z}}+\frac{\phi}{L_{z}},
\quad\varepsilon_{-}(q,\vec{k}_{\perp}=0)/\upsilon=-q=-\frac{2n\pi}{L_{z}}-\frac{\phi}{L_{z}}.
\end{eqnarray}
Here the point is that the relative energy eigenvalues depend only on one parameter $\phi$.
Therefore the accidental band crossing is possible when both $\varepsilon_{+}$
and $\varepsilon_{-}$ become zero simultaneously, i.e., $\phi=-2n\pi$.
Since $n$ can take any integer value, successive gap closing is possible by
tuning one parameter $\phi$ continuously.

On the other hand, the gap-closing condition under the open boundary condition
is more complicated. After some straightforward calculation as shown in detail below, it can be shown that
in general the energy eigenvalue of a single Weyl point can be written as
\begin{eqnarray}
\varepsilon_{+}(q,\vec{k}_{\perp}=0)/\upsilon=\frac{(2n+1)\pi}{2L_{z}}+\alpha,
\quad\varepsilon_{-}(q,\vec{k}_{\perp}=0)/\upsilon=-\frac{(2n+1)\pi}{2L_{z}}+\alpha,
\end{eqnarray}
where $\alpha$ is a constant.
Note that the relative energy between $\varepsilon_{+}$ and $\varepsilon_{-}$ does not depend on $\alpha$.
Therefore the variation of the parameter $\alpha$ just induces the overall shift of the full energy spectrum
and cannot induce band crossings at a single Weyl point.
However, when two Weyl points are coupled, accidental band crossing is possible.
For example, when a pair of Weyl points are along
the surface normal direction, the open boundary condition induces mixing between these two Weyl points.
Because of this mixing, the boundary condition induces a strong constraint to possible energy eigenvalues,
which are given by
\begin{eqnarray}
\varepsilon_{1,+}(q,\vec{k}_{\perp}=0)/\upsilon=\frac{n\pi}{L_{z}}+\alpha+\beta,
\quad\varepsilon_{1,-}(q,\vec{k}_{\perp}=0)/\upsilon=-\frac{n\pi}{L_{z}}+\alpha+\beta,
\nonumber\\
\varepsilon_{2,+}(q,\vec{k}_{\perp}=0)/\upsilon=\frac{n\pi}{L_{z}}-\alpha+\beta,
\quad\varepsilon_{2,-}(q,\vec{k}_{\perp}=0)/\upsilon=-\frac{n\pi}{L_{z}}-\alpha+\beta,
\end{eqnarray}
where $\alpha$, $\beta$ are constants.
It is worth to note that
while the energy levels for a Weyl point i.e., either $\varepsilon_{1,\pm}$ or $\varepsilon_{2,\pm}$
are shifted in parallel,
the relative energy between $\varepsilon_{1,\pm}$ and $\varepsilon_{2,\pm}$
depends on one parameter $\alpha$.
Since $\beta$ describes the overall shift of the energy,
it can be absorbed to the chemical potential of the system, hence can be neglected. (Let $\beta=0$.)
Then, for example, if $\alpha=n\pi/L_{z}$, $\varepsilon_{1,-}$=$\varepsilon_{2,+}$=0
is possible for any integer $n$.
Therefore successive gap-closing can be achieved by changing the parameter $\alpha$ continuously.

Now let us describe the detailed derivation of the gap-closing condition
under the open boundary condition. Here the idea is to understand
how the energy spectrum near a Weyl point depends on hard wall
boundary conditions. In general, a pair of Weyl fermions with opposite chiral charge
can be described by
\begin{displaymath}
-i\upsilon\vec{\alpha}\cdot\nabla\psi=E\psi,\quad
\vec{\alpha}=
\left(\begin{array}{cc}
\vec{\sigma} & 0\\
0 & -\vec{\sigma}
\end{array}\right)=\tau_{3}\vec{\sigma},
\end{displaymath}
where the Pauli matrices $\tau_{1,2,3}$ describe two Weyl fermions and
the velocity $\upsilon$ is assumed to be isotropic.
Following McCann and Fal'ko,~\cite{McCann} we introduce
a confinement potential at the boundary $\vec{r}=\vec{r}_{B}$
to describe a hard wall boundary condition for Weyl fermions
in the following way,
\begin{eqnarray}
[-i\upsilon\vec{\alpha}\cdot\nabla+c\upsilon A \delta(\vec{r}-\vec{r}_{B})]\psi=E\psi,
\end{eqnarray}
where $c$ is a real constant and $A$ is an arbitrary 4$\times$4 Hermitian, unitary matrix
satisfying $AA^{\dag}=A^{2}=1$.
We assume that $\hat{n}_{B}$ is the surface normal direction of the hard wall
and the wave function $\psi$ becomes zero outside the film. Then after integrating
over a small region near the boundary, we obtain the following constraint.
\begin{eqnarray}
-i\vec{\alpha}\cdot\hat{n}_{B}\psi(\vec{r}_{B})=cA\psi(\vec{r}_{B}).
\end{eqnarray}
From the equation above, it is straightforward to derive the following relations,
\begin{eqnarray}
c=1,\quad \{\vec{\alpha}\cdot\hat{n}_{B},A\}=0,
\quad \psi(\vec{r}_{B})=i\vec{\alpha}\cdot\hat{n}_{B}A\psi(\vec{r}_{B})\equiv M\psi(\vec{r}_{B}).
\end{eqnarray}
Therefore the boundary condition can be represented by the following local linear
constraint to the wave function
\begin{eqnarray}
\psi(\vec{r}_{B})=M\psi(\vec{r}_{B})
\end{eqnarray}
in which $M$ satisfies
\begin{eqnarray}\label{eqn:M}
M=M^{\dag},\quad M^{2}=1,\quad \{\hat{n}_{B}\cdot\vec{J},M\}=0,
\end{eqnarray}
where $\vec{J}=\partial_{\vec{k}}H$.
Here the anticommutation relation $\{\hat{n}_{B}\cdot\vec{J},M\}=0$
implies the absence of current along the surface normal direction.~\cite{Beenakker}

The next step is to find the matrix $M$ satisfying Eq.~(\ref{eqn:M}).
The most general form of $M$ can be written as
\begin{eqnarray}
M=\sum_{i,j=0}^{3} c_{ij}(\tau_{i}\otimes \sigma_{j}),\quad c_{ij}=\text{real},
\end{eqnarray}
Using $\vec{J}=\upsilon\tau_{z}\vec{\sigma}$ and imposing the constraints in Eq.~(\ref{eqn:M}),
we can show that the most general form of $M$ is given by $M=M_{1}+M_{2}$ in which
\begin{eqnarray}\label{eqn:M1}
M_{1}=m_{1}\cos\theta\tau_{x}\sigma_{0}+m_{1}\sin\theta\tau_{y}(\hat{n}_{B}\cdot\vec{\sigma})
-m_{2}\sin\theta\tau_{x}(\hat{n}_{B}\cdot\vec{\sigma})+m_{2}\cos\theta\tau_{y}\sigma_{0},
\end{eqnarray}
and
\begin{eqnarray}\label{eqn:M2}
M_{2}=\tau_{z}(\vec{n}_{1}\cdot\vec{\sigma})+\tau_{0}(\vec{n}_{2}\cdot\vec{\sigma}),
\end{eqnarray}
where $\vec{n}_{1}=m_{3}\cos\theta\hat{n}_{1}$ and $\vec{n}_{2}=m_{3}\sin\theta\hat{n}_{2}$.
Here $m_{3}$ and the unit vectors $\hat{n}_{1}$ and $\hat{n}_{2}$ satisfy
$m_{1}^{2}+m_{2}^{2}+m_{3}^{2}=1$, $\hat{n}_{1}\cdot\hat{n}_{2}=\hat{n}_{1}\cdot\hat{n}_{B}=\hat{n}_{2}\cdot\hat{n}_{B}=0$.
Therefore, in general, $M$ can be characterized by 4 independent parameters,
$m_{1}$, $m_{2}$, $\theta$, and another angular variable describing
the orientation of $\hat{n}_{1}$ or $\hat{n}_{2}$.
Moreover, from the structure of $M_{1,2}$,
we can easily see that the Weyl fermion pair
are coupled (decoupled) under $M_{1}$ ($M_{2}$).
Therefore when we describe
the Weyl fermion pairs of a [111] film, which are not along the surface normal direction,
since they are not coupled under the open boundary condition
due to the translational invariance within the plane parallel to the surface,
the boundary condition should be described by $M=M_{2}$.
On the other hand, when the pair of Weyl points are coupled under the boundary condition, in general,
$M=M_{1}+M_{2}$ is possible. However,
in the case of the Weyl fermion pair along [111] direction,
additional rotational symmetry with respect to [111] direction (or $\hat{n}_{B}$) can be imposed.
Hence $M_{2}$ term is not allowed and $M=M_{1}$ should be satisfied.

Let us first discuss the case of decoupled Weyl fermion pairs with $M=M_{2}$.
If we assume that $\hat{n}_{B}=\mu\hat{z}$ with $\mu=\pm$,
$\vec{n}_{1,2}$ in Eq.~(\ref{eqn:M2})
can be written as
\begin{eqnarray}
\vec{n}_{1}=\mu\cos\theta_{\mu} (\cos\xi_{\mu},-\sin\xi_{\mu},0),
\quad \vec{n}_{2}=\mu\sin\theta_{\mu} (\sin\xi_{\mu},\cos\xi_{\mu},0).
\end{eqnarray}
Here $\mu=+$ ($\mu=-$) describes the top (bottom) surface at $z=+L_{z}/2$ ($z=-L_{z}/2$).
The wave function for two Weyl fermions can be written as $\psi^{T}=(\psi_{1}^{T},\psi_{2}^{T})$
where
\begin{displaymath}
\psi_{1}(\vec{r})=Ae^{iqz}e^{i\vec{k}_{\perp}\cdot\vec{r}_{\perp}}
\left(\begin{array}{c}
k_{x}-ik_{y}\\
\varepsilon/\upsilon-q
\end{array}\right)+
Be^{-iqz}e^{i\vec{k}_{\perp}\cdot\vec{r}_{\perp}}
\left(\begin{array}{c}
k_{x}-ik_{y}\\
\varepsilon/\upsilon+q
\end{array}\right),
\end{displaymath}
and
\begin{displaymath}
\psi_{2}(\vec{r})=Ce^{iqz}e^{i\vec{k}_{\perp}\cdot\vec{r}_{\perp}}
\left(\begin{array}{c}
-k_{x}+ik_{y}\\
\varepsilon/\upsilon+q
\end{array}\right)+
De^{-iqz}e^{i\vec{k}_{\perp}\cdot\vec{r}_{\perp}}
\left(\begin{array}{c}
-k_{x}+ik_{y}\\
\varepsilon/\upsilon-q
\end{array}\right),
\end{displaymath}
The discrete $q$ value can be determined from the condition of $\psi=M\psi$
at $z=\pm L_{z}/2$. Since we are interested in
the gap-closing condition at the Weyl point, we can assume $k_{x}=k_{y}=0$.
Then using the discrete $q$ value, the corresponding energy eigenvalue
$\varepsilon=\upsilon s q$ ($s=\pm$) can be written as
\begin{eqnarray}\label{eqn:decoupled1}
\varepsilon_{1}=\upsilon s q=\upsilon \Big[s\frac{(2n_{1}+1)}{2L_{z}}\pi
-\frac{(\theta_{+}-\theta_{-})}{2L_{z}}+\frac{(\xi_{+}-\xi_{-})}{2L_{z}}\Big],
\end{eqnarray}
for one Weyl fermion. Similarly, for the other Weyl fermion,
\begin{eqnarray}\label{eqn:decoupled2}
\varepsilon_{2}=\upsilon s q=\upsilon \Big[s\frac{(2n_{2}+1)}{2L_{z}}\pi
-\frac{(\theta_{+}-\theta_{-})}{2L_{z}}-\frac{(\xi_{+}-\xi_{-})}{2L_{z}}\Big].
\end{eqnarray}
Here we can see that the energy eigenvalues explicitly depend
on several parameters coming from the boundary condition.
However, the energy spectrum
is only globally shifted in each Weyl point and the relative energy between
discrete eigenstates remains the same.
It is to note that an accidental gap-closing is possible only when
the energy gap between the conduction band and valence band
can be tuned to zero. Therefore under the open boundary condition,
two decoupled Weyl point cannot show an accidental gap closing.

Now let us consider the case of two coupled Weyl points. In this
case the boundary condition can be imposed by using $M=M_{1}$.
The most general form of $M_{1}$ can be written as
\begin{eqnarray}
M_{1}=\cos\phi\cos\theta\tau_{x}\sigma_{0}+\cos\phi\sin\theta\tau_{y}(\hat{n}_{B}\cdot\vec{\sigma})
-\sin\phi\sin\theta\tau_{x}(\hat{n}_{B}\cdot\vec{\sigma})+\sin\phi\cos\theta\tau_{y}\sigma_{0},
\end{eqnarray}
which can be parametrized by two angular variables $\theta$ and $\phi$.
After repeating similar calculation, the energy spectrum
at $k_{x}=k_{y}=0$ can be obtained as
\begin{eqnarray}\label{eqn:coupled1}
\varepsilon_{1}=\upsilon s q=\upsilon \Big[s\frac{n_{1}\pi}{L_{z}}
-\frac{(\theta_{+}+\theta_{-})}{2L_{z}}+s\frac{(\phi_{+}+\phi_{-})}{2L_{z}}\Big],
\end{eqnarray}
and
\begin{eqnarray}\label{eqn:coupled2}
\varepsilon_{2}=\upsilon s q=\upsilon \Big[s\frac{n_{2}\pi}{L_{z}}
-\frac{(\theta_{+}+\theta_{-})}{2L_{z}}-s\frac{(\phi_{+}+\phi_{-})}{2L_{z}}\Big],
\end{eqnarray}
Notice that contrary to the case of the decoupled Weyl points
in Eq.~(\ref{eqn:decoupled1})and Eq.~(\ref{eqn:decoupled2}),
the relative energy between discrete energy eigenvalues depends on
parameters resulting from the boundary condition.
In this case, accidental band crossings can be achieved as explained
in the beginning of this section.

\bibliographystyle{naturemag}


{\small \subsection*{Acknowledgements}
We are grateful for support from the Japan Society for the Promotion of Science (JSPS) through the `Funding Program for World-Leading Innovative R\&D on Science and Technology (FIRST Program). We thank David Vanderbilt for his insightful comments.}






\newpage
\noindent{\bf Supplementary Information}
\\
{\bf Localized states due to lattice geometry in the kagome lattice.}

One peculiar property of [111] films is that
there are additional non-topological surface states localized on the surface
terminated by a triangular lattice. Here we will show that
those additional localized states originate from
the geometrical structure of the pyrochlore lattice.

Before we discuss about the localized states on the pyrochlore lattice, let us first
consider the case of the kagome lattice which also
supports non-topological localized states on the edge.
The lattice structure of a kagome lattice is shown in Fig. 8c.
To describe surface states we introduce a kagome ribbon indicated
by the shaded region in Fig. 8c.
A kagome ribbon has a sawtooth shape (a straight line shape) on the top (bottom)
edge while the translational invariance is maintained along the horizontal direction ($x$-direction).
One simple way to make a kagome ribbon
is vertical stacking of one-dimensional chains,
each of which is composed of up-triangles connected along the $x$-direction.
Similar to the case of [111] films composed of stacked bilayers,
the two edges of a kagome ribbon have asymmetric structures.

Now let us consider a tight-binding Hamiltonian describing hopping processes
between nearest neighbor sites, which can be written as
\begin{eqnarray}
H=t\sum_{\langle i,j\rangle}[c^{\dag}_{i}c_{j}+c^{\dag}_{j}c_{i}],
\end{eqnarray}
where $i$ and $j$ are indices denoting the kagome lattice sites
and $\langle i,j\rangle$ indicates a nearest neighbor pair of sites.
Fig. S1a shows the
band structure of a kagome ribbon
under the periodic boundary condition along the vertical direction.
One interesting property of the band structure is that
there are many flat bands at the energy $E=-2t$.
Such a flat band appears due to the localized state
existing in a hexagonal plaquette. (See Ref. 32.) Explicitly,
the wave function of the state localized in a hexagon can be written as
\begin{eqnarray}
|\Psi_{\text{Hex}}\rangle=\frac{1}{\sqrt{6}}\sum_{n=1}^{6}(-1)^{n}|n\rangle,
\end{eqnarray}
where $|n\rangle\equiv c^{\dag}_{n}|0\rangle$ ($n$=1,...,6)
indicates the six sites around a hexagon.
Here $|0\rangle$ indicates the vacuum state.
An example of such a localized state in a hexagon is marked by red dots in Fig. 8c,
in which $\pm$ indicates the sign in front of the state $|n\rangle$ constituting
$|\Psi_{\text{Hex}}\rangle$.
Since the number of hexagons is equal to the number of unit cells,
a flat band can be formed by taking a linear superposition of
every localized state in a hexagon. In fact, under the periodic
boundary condition, localized hexagon states are under a constraint
such that the summation of all localized hexagon states should vanish
due to the absence of boundaries in a periodic system.
However, since there are two additional degenerate localized states,
which are so-called non-contractible loop states,
a flat band can spread over the full Brillouin zone
with additional band touching at the $\Gamma$ point. (See Ref. 32.)

Under the open boundary condition, the tight-binding Hamiltonian
on a kagome ribbon
shows an unexpected interesting feature.
Fig. S1b describes the band structure of the system
under the open boundary condition.
Here we can see the appearance of an additional band, marked by a blue line in Fig. 8b,
between the flat band and its neighboring dispersive band.
By counting the number of flat bands at a fixed momentum,
we can see that the additional blue band is a part
of the flat band under the periodic boundary condition.
Surprisingly, this additional band is exponentially localized
near the sawtooth shaped edge of the strip as shown in Fig. 8d.
The localization length of this additional state
is inversely proportional to the energy gap between
the flat band and its neighboring dispersive band at the given momentum.
Let us call this localized state as a geometry-induced surface state (GSS).
Considering the fact that the origin of the flat band is the localized state
in a hexagon, we can easily find the origin of the GSS.
As shown in Fig. 8c, at the sawtooth shaped edge,
there are a series of broken hexagons aligned along the edge.
It is to be noted that it is inevitable to create broken hexagons
once a sawtooth shaped edge is introduced.
The localized state related with a broken hexagon should be a part of
flat bands under the periodic boundary condition. However,
once a sawtooth shaped edge is introduced, the flat band
which contains the localized states in hexagons touching the edge splits from
the other flat bands, and forms the GSS.

To understand the relationship between the broken hexagon states and the GSS
more explicitly,
we can consider the following superposition of broken hexagon states,
\begin{eqnarray}
|\Psi\rangle=\sum_{\ell}A_{\ell}\{-x|\ell,1\rangle+|\ell,2\rangle-|\ell,3\rangle\},
\end{eqnarray}
where the index $\ell$ indicates a broken hexagon and $|\ell,i\rangle$ ($i=1,2,3$)
indicates the three sites in a corresponding broken hexagon.
As can be seen in Fig. S1c, $|\ell,4\rangle=|\ell+1,1\rangle$.
The coefficients in front of $|\ell,2\rangle$ and $|\ell,3\rangle$ are fixed
to find a state completely localized on the edge.
After some straightforward calculations, it can be shown that
the wave function of GSS at the momentum $k=0$ and $k=\pi$
can be written as
\begin{eqnarray}
|\Psi_{\text{GSS}}(k=0)\rangle=N_{0}\sum_{\ell}\{|\ell,2\rangle-|\ell,3\rangle\},
\end{eqnarray}
and
\begin{eqnarray}
|\Psi_{\text{GSS}}(k=\pi)\rangle=N_{\pi}\sum_{\ell}(-1)^{\ell}\{-\sqrt{2}|\ell,1\rangle+|\ell,2\rangle-|\ell,3\rangle\},
\end{eqnarray}
where $N_{0,\pi}$ are normalization constants.
It means that the GSS is completely localized within broken hexagons at $k=\pi$.
As we move away from $k=\pi$, $\Psi_{\text{GSS}}(k)$ shows small spreading
along the vertical direction but is still exponentially localized
near the sawtooth edge as shown in Fig. 8d.
Since its localization length is determined by the gap between the flat band
and its neighboring dispersive bulk bands, the GSS has stability.
If additional hopping processes are considered, the localization length
of the GSS can be changed due to the modified bulk band structure.
However, as long as there is a finite gap at a given momentum,
the GSS can be exponentially localized on the sawtooth edge.

In the case of $\Psi_{\text{GSS}}(k=0)$, it is energetically degenerate with
other flat bands coming from the localized states in hexagons.
$\Psi_{\text{GSS}}(k=0)$ has nonzero amplitude only on site 2 and 3
in a broken hexagon and such a structure appears identically in every broken hexagon.
In fact, $\Psi_{\text{GSS}}(k=0)$ is nothing but one of the non-contractible loop states,
which extends along the $x$-direction parallel to the sawtooth edge.
In contrast to the case of 2D periodic systems where there are two non-contractible loop states,
under the open BC
there is only one non-contractible loop states since the geometry
of the strip is basically a cylinder.

{\bf Localized states due to lattice geometry in [111] films.}
The additional localized states existing in [111] films can be
understood by extending the idea of GSS in the kagome lattice
to the pyrochlore lattice in a straightforward way.
Pyrochlore lattice is composed of kagome planes
which are stacked along $[111]$ or its three other symmetry equivalent directions.
Let us first consider the tight-binding Hamiltonian containing only
the nearest neighbor hopping amplitudes.
Since each kagome plane supports localized states in each hexagon,
the total number of localized states in hexagons is four times larger than
the total number of unit cells in a pyrochlore lattice.
However, those localized states in hexagons are not linearly independent.
As discussed by Bergman et al. in Ref. 32 in the main text, if we consider a volume enclosed by
four neighboring hexagons, the superposition
of corresponding four localized states can be vanished.
Taking into account of such local constraints, the total number
of linearly independent localized states is almost two times
larger than the number of unit cells, which is consistent
with the fact that there are two completely flat bands
in the bulk band structure as shown in Fig. 9b.

In the case of a [111] film, there are kagome planes
stacked along the surface normal direction ($z$-direction) of the film.
Because of the translational invariance in $xy$ planes,
those kagome planes do not have broken hexagons.
On the other hand, we also have to consider another group of
parallel kagome planes to fully account for the flat bands
in the bulk band structure. The surface normal direction
of this second group of kagome planes is not parallel to
the $z$-direction. One example of such a slanted kagome
plane is shown in Fig. 9a.
Here the point is because of the structure of the film,
each slanted kagome plane should have broken hexagons on the top
surface of the film terminated by a triangular lattice plane.
Such broken hexagons can generate GSS which is exponentially localized
on the top surface of the film. To prove that
the GSS results from the states in broken hexagons,
we compute the energy spectrum of the film which has
the triangular (kagome) lattice on the top (bottom)
surface as shown in Fig. 9c.
By comparing Fig. 9b and Fig. 9c,
we can clearly see the emergence of the surface state (GSS)
in a gap between the flat band and its neighboring dispersive bulk band,
which is marked by a blue line in Fig. 9c.
Also by counting the number of flat bands in both cases,
we can see that the GSS is derived from the flat bands in Fig. 9b.
Hence, the GSS is mainly from the states in broken hexagons on the top surface.
Moreover, since the GSS is exponentially localized near the top surface
and its localization length is given by the gap between the flat band and its neighboring
dispersive bulk band, the GSS is stable against external perturbation
as long as the gap remains finite.
In Fig. 9d and Fig. 9e, we compare
the band structure of the tight-binding Hamiltonian
including the second nearest hopping and various effective spin dependent
hopping processes.
Due to the additional hopping processes, all bands are basically dispersive.
But here we can still observe the existence of the GSS clearly.
When the system possess another surface states such as the Fermi arc states
of a Weyl-SM, the coupling between the Fermi arc states and the GSS can induce
nontrivial topological phenomena driven by these surface states,
which is the main topic discussed in the main text of the paper.

\

\newpage




\begin{table*}[h]
\begin{tabular}{@{}|c|c|c|}
\hline
\hline
& Top Kagome Layer& Top Triangular Layer\\
& (Topological-SS only)& (Topological-SS + Geometrical-SS) \\
\hline
Bottom Kagome Layer& No AHE & Giant AHE\\
(Topological-SS only)& (Metallic states in the gap) & (Metallic states in the gap)\\
\hline
Bottom Triangular Layer& Giant AHE & No AHE  \\
(Topological-SS + Geometrical-SS)& (Metallic states in the gap)& (No metallic states in the gap)\\
\hline \hline
\end{tabular}
\end{table*}
{\noindent {\bf Table 1:}
{\bf Topological properties of [111] thin films for various surface termination
when the bulk is a fully gapped insulator
with $m_{1}<m<m_{2}$.}
}

\newpage

\begin{figure*}[t]
\centering
\includegraphics[width=16 cm]{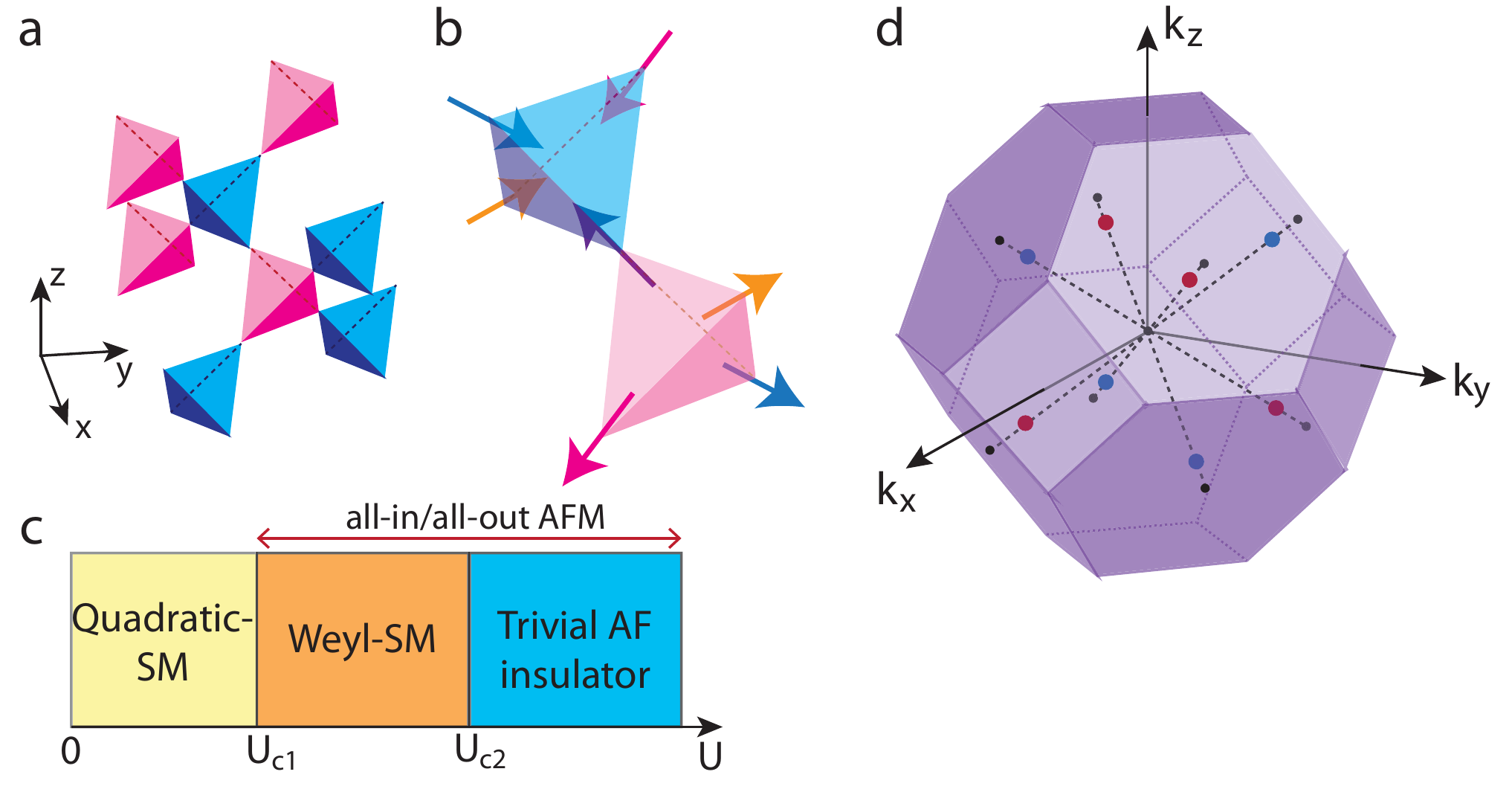}
\caption{
{\bf Lattice structure and phase diagram of the 3D bulk system.}
({\bf a}) Lattice structure of the pyrochlore lattice composed of
corner sharing tetrahedrons.
({\bf b}) Local spin structure of the AIAO antiferromagnet (AFM).
({\bf c}) Schematic 3D bulk phase diagram as a function of local Coulomb interaction $U$.
For $U>U_{c1}$, the system is an AIAO antiferromagnetic (AF) state.
({\bf d}) Distribution of eight Weyl points for the Weyl-SM phase.
All eight Weyl points are aligned along the [111] or its symmetry equivalent directions.
Here red (blue) dot indicates a Weyl point with the chiral charge +1 (-1).
} \label{fig:3Dlattice}
\end{figure*}

\begin{figure*}[t]
\centering
\includegraphics[width=16 cm]{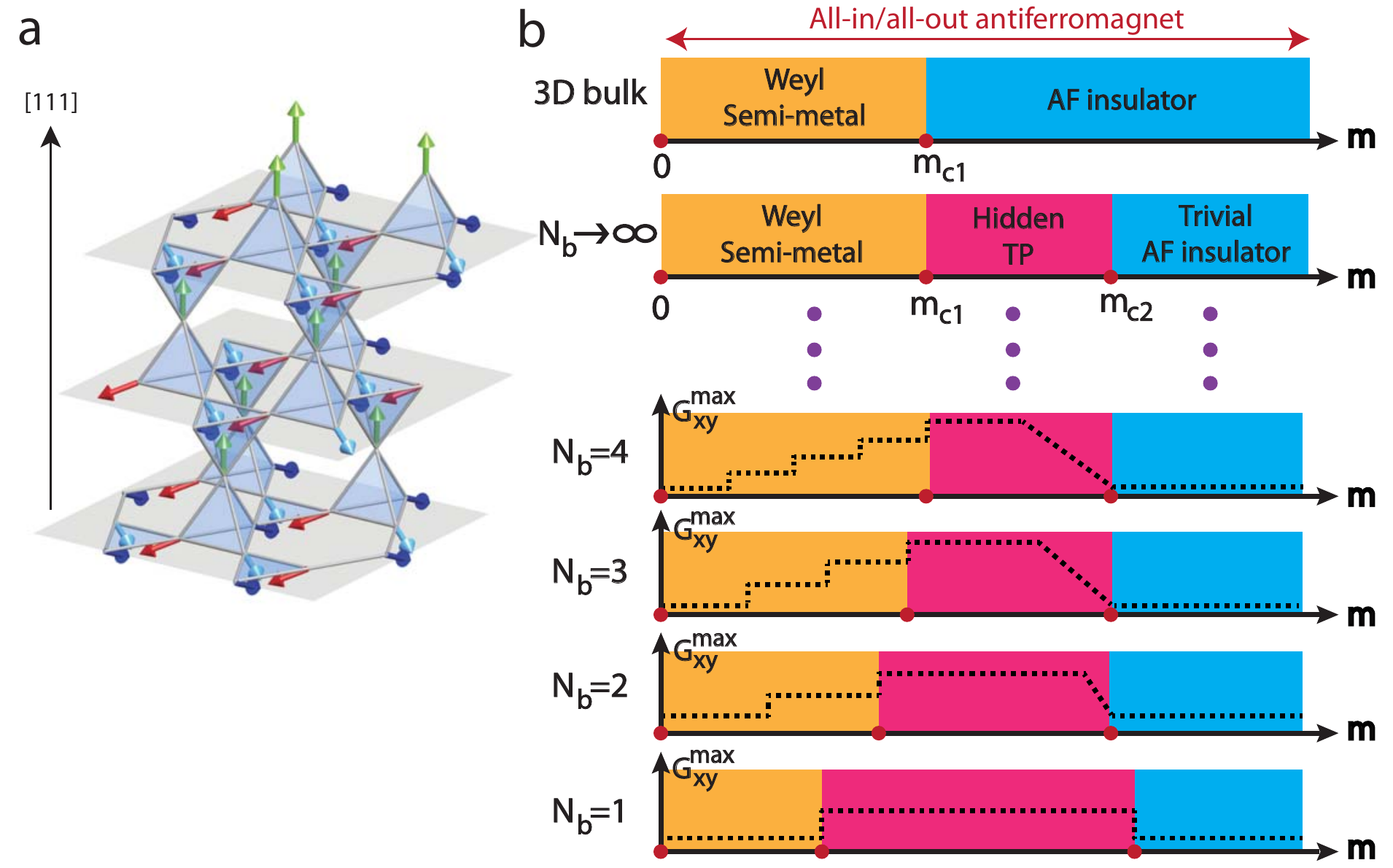}
\caption{
{\bf Dimensional crossover for [111] films of pyrochlore iridate antiferromagnets.}
({\bf a}) Lattice structure and spin ordering pattern of the [111] film with AIAO spin configuration.
Along the [111] direction, the kagome and triangular lattices are stacked alternatively.
Neighboring kagome and triangular layers constitute a unit bilayer.
({\bf b}) Evolution of the phase diagram of [111] thin films when the thickness of films increases.
Here $m=\langle S_{i}\rangle$ indicates the average local spin moment of the AIAO antiferromagnetic (AF)
ground state and $N_{b}$ denotes the number
of bilayers. The dotted lines in the phase diagram indicate the change
of the maximum anomalous Hall conductance $G^{\text{max}}_{xy}$ of the system.
The detailed changes of $G_{xy}$ and $G^{\text{max}}_{xy}$ are shown in Fig. 3b.
The pink region indicates the phase where the bulk states are fully gapped
but the Hall conductance of the system is nonzero, which is derived from
the hidden topological phase (TP) in the bulk limit.
} \label{fig:dimensionalcrossover}
\end{figure*}

\begin{figure*}[t]
\centering
\includegraphics[width=16 cm]{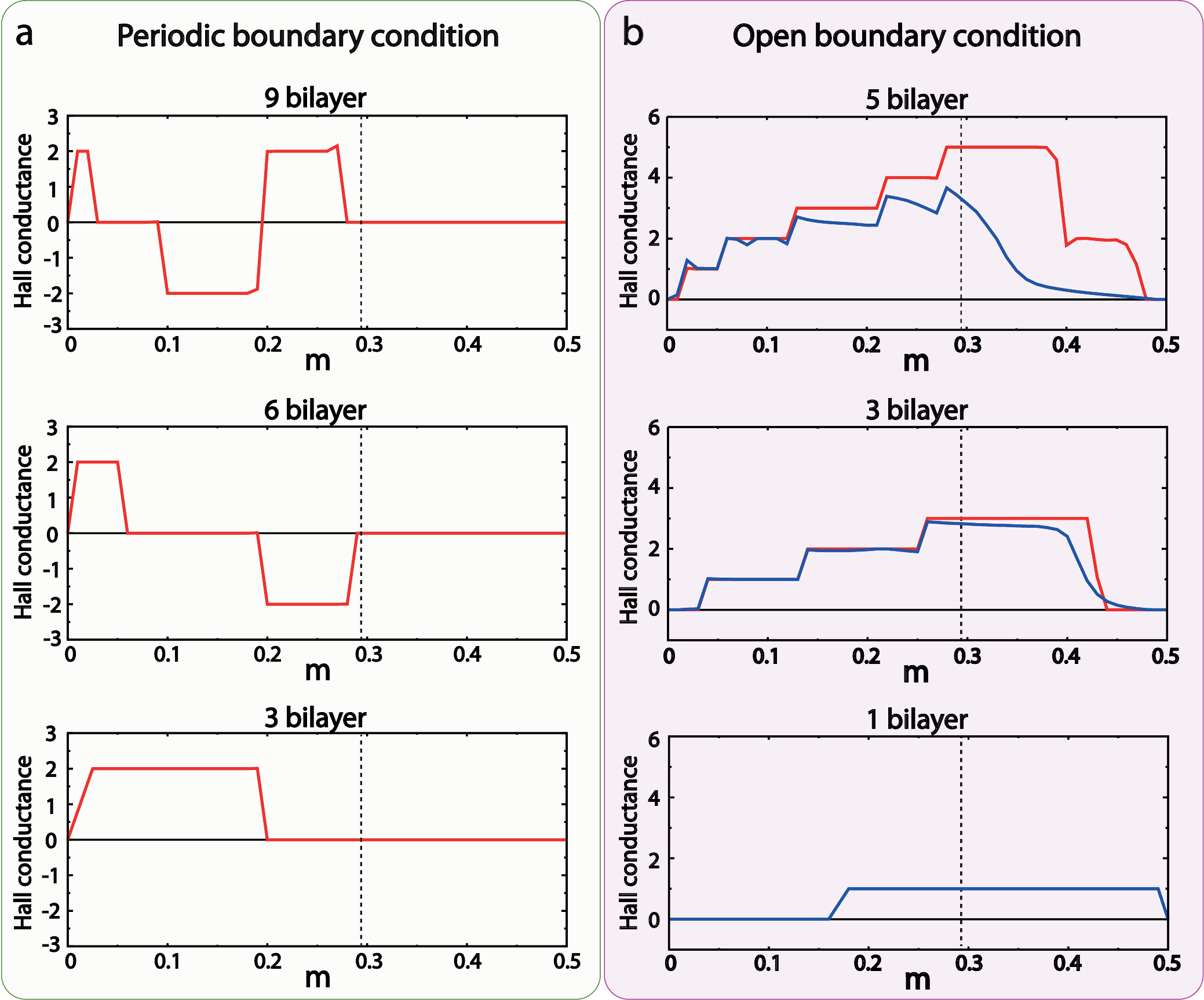}
\caption{
{\bf Variation of the anomalous Hall conductance $G_{xy}$ of [111] films under the periodic and open
boundary conditions as a function of $m$. }
({\bf a}) Under the periodic boundary condition. Oscillating $G_{xy}$ appears.
Under the periodic boundary condition $G_{xy}=G^{\text{max}}_{xy}$
because there is no surface state in the bulk energy gap.
For $m>m_{c1}=0.29$, the anomalous Hall conductance becomes zero.
({\bf b}) Under the open boundary condition.
$G^{\text{max}}_{xy}$ (the red line) increases monotonously
until it reaches the maximum value of $G^{\text{max}}_{xy}=N_{b}e^{2}/h$ near $m=m_{c1}$.
For $N_{b}=1$, $G_{xy}$ and $G^{\text{max}}_{xy}$ are the same.
$G_{xy}$ (the blue line) basically follows the behavior of $G^{\text{max}}_{xy}$ in the case of thin films with $N_{b}\leq 4$.
When $N_{b}\geq 5$, the surface states in the bulk gap disturb the monotonous increment of $G_{xy}$.
The vertical dotted line indicates the critical point $m_{c1}$.
} \label{fig:boundarycondition}
\end{figure*}

\begin{figure*}[t]
\centering
\includegraphics[width=16 cm]{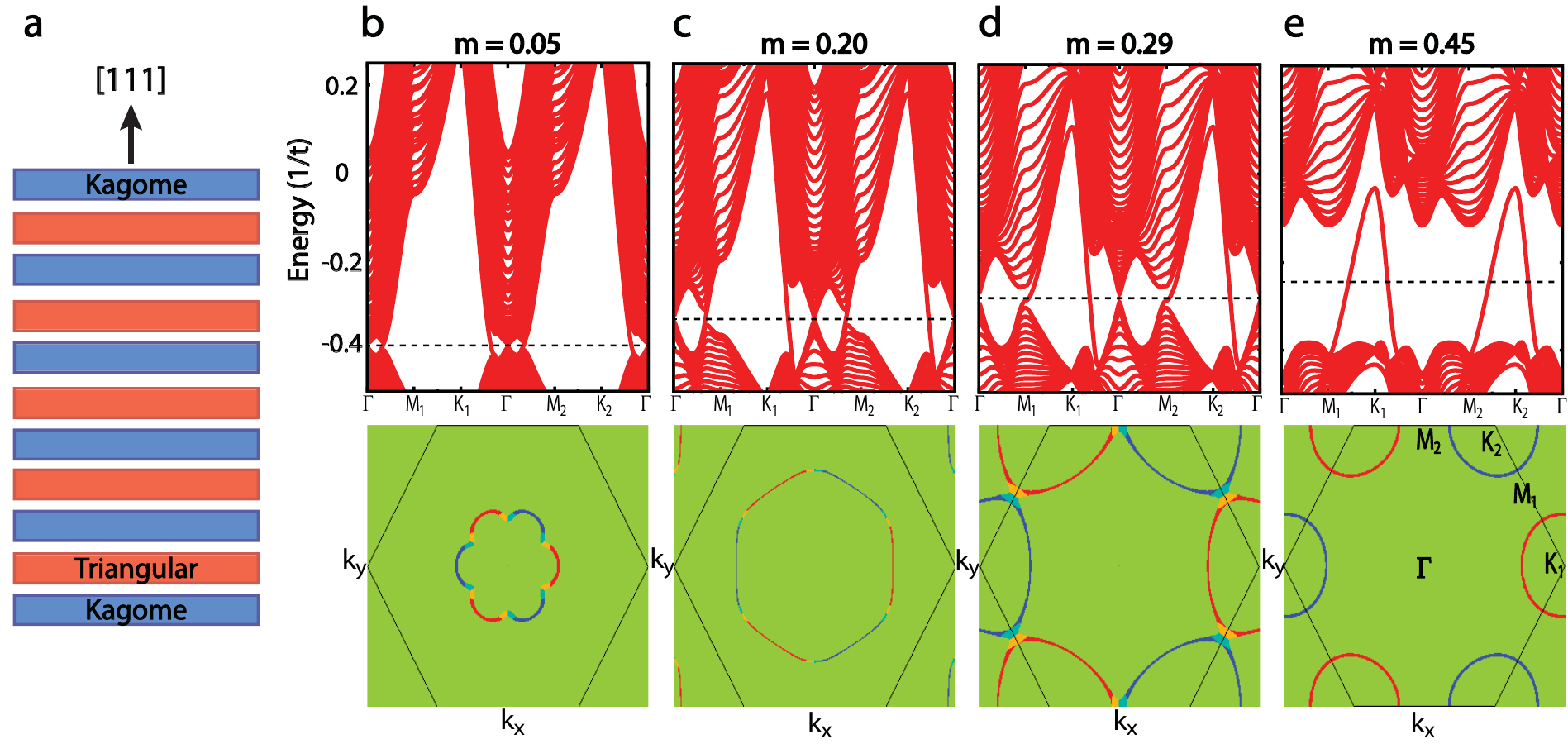}
\caption{
{\bf Evolution of the band structure and surface spectrum of [111] films
which are terminated with the kagome lattices on both surfaces of the film.}
(a) Structure of the thin film having the kagome lattice on both the top and bottom layers.
Here the film is composed of $N_{b}$=20 bilayers, plus an additional Kagome layer on the top.
(b-e) The band structure near the Fermi energy $E_{F}$ (upper panels) and corresponding
Fermi surface shape due to localized in-gap states (lower panels).
({\bf b}) for $m=0.05$ (Weyl-SM),
({\bf c}) for $m=0.20$ (Weyl-SM), ({\bf d}) for $m=m_{c1}=0.29$ (critical point),
and ({\bf e}) for $m=0.45$ (gapped bulk insulator with surface metallic states).
Here the dotted lines on top panels indicate the Fermi energy $E_{F}$.
In the lower panels, the red (blue) lines indicate the states at $E_{F}$
localized on the top (bottom) surfaces and there is no surface state in the green region.
When $m<m_{c1}$, the size of the Fermi surface increases as $m$ grows (({\bf b}) and ({\bf c})).
At $m=m_{c1}$, the Fermi surface shows Lifshitz transition (({\bf d})).
Finally, when $m>m_{c1}$, the Fermi surface is composed
of two isolated Fermi loops encircling two corners of the surface
Brillouin zone (({\bf e})).
} \label{fig:KKfilm}
\end{figure*}

\begin{figure*}[t]
\centering
\includegraphics[width=16 cm]{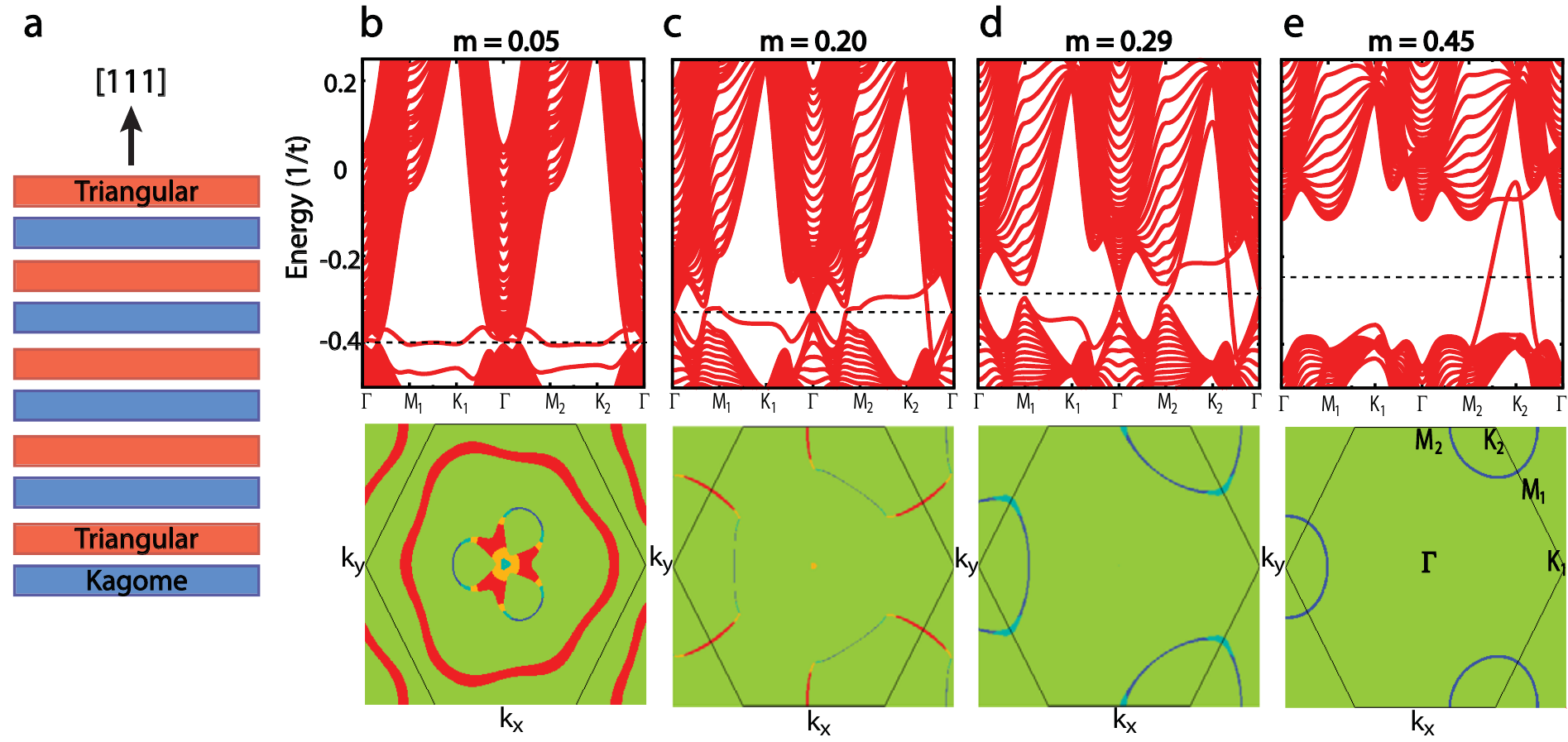}
\caption{
{\bf Evolution of the band structure and surface spectrum of films ($N_{b}$=20)
which have the triangular (kagome) lattices on the top (bottom) surface.}
(a) Structure of the thin film having the triangular (kagome) lattice on the top (bottom) layer.
(b-e) The band structure near the Fermi energy $E_{F}$ (upper panels) and corresponding
Fermi surface shape due to localized in-gap states (lower panels).
({\bf b}) for $m=0.05$ (Weyl-SM),
({\bf c}) for $m=0.20$ (Weyl-SM), ({\bf d}) for $m=m_{c1}=0.29$ (critical point),
and ({\bf e}) for $m=0.45<m_{c2}$ (hidden topological phase).
Here the dotted lines on top panels indicate the Fermi energy $E_{F}$.
In the lower panels, the red (blue) lines indicate the states
localized on the top (bottom) surfaces and there is no surface state in the green region.
When $m<m_{c1}$ (({\bf b}), ({\bf c})), additional non-topological states are hybridized with Fermi arc states
on the top layer while the Fermi arc states stably survive on the bottom layer.
When $m\geq m_{c1}$ (({\bf d}), ({\bf e})), only the states localized on the bottom layer
touch the Fermi level. Finally, when $m>m_{c2}$,
the states on the top and bottom surfaces decouple, leading to the fully gapped
trivial AFI.
} \label{fig:KTfilm}
\end{figure*}
\begin{figure*}[t]
\centering
\includegraphics[width=16 cm]{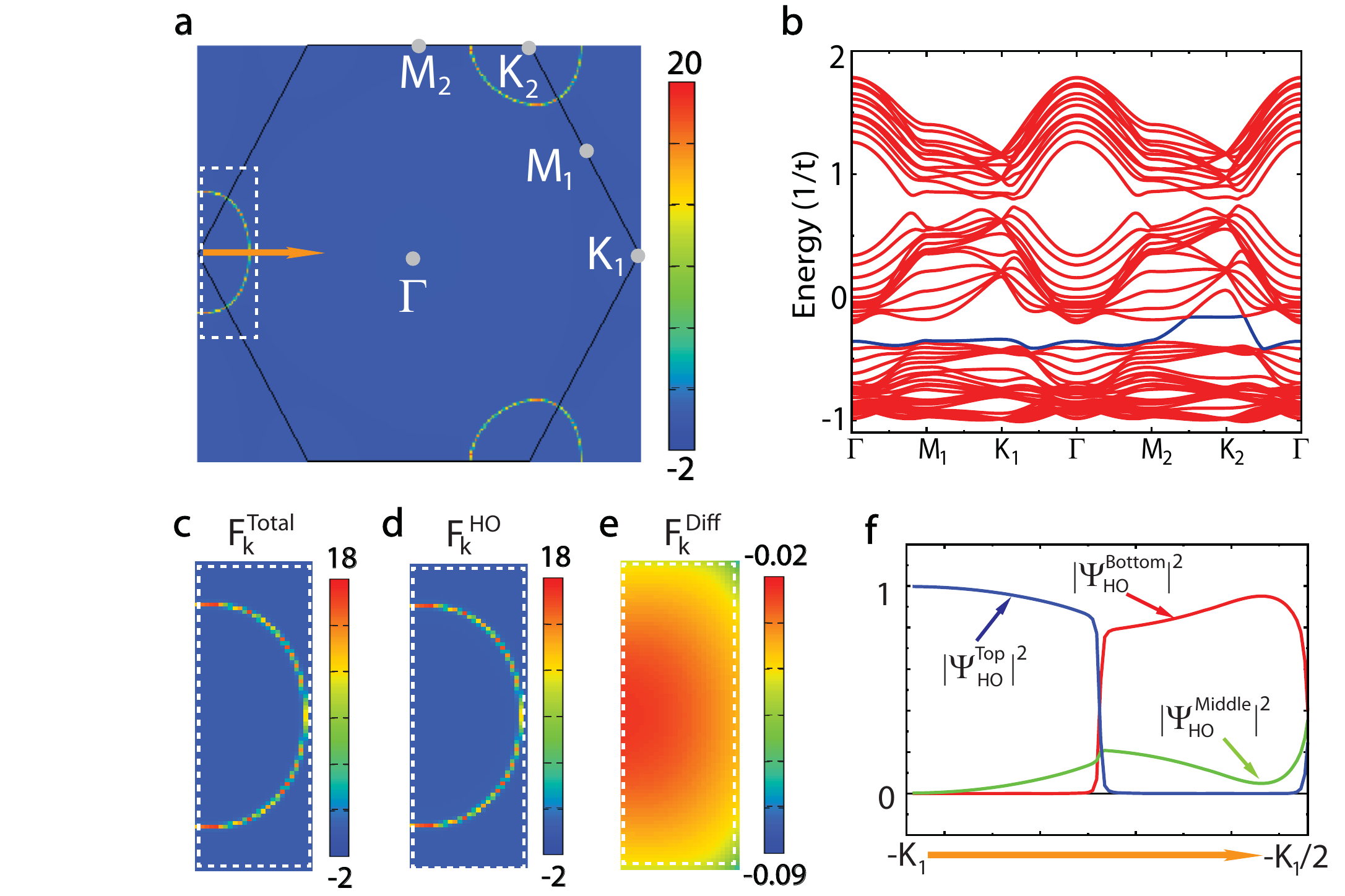}
\caption{
{\bf Band crossing between the top and bottom surface states and the resulting
Berry curvature distribution for a film composed of 6 bilayers.}
({\bf a}) Distribution of the Berry curvature
of the occupied bands, $F^{\text{Total}}_{\textbf{k}}$ in the surface Brillouin zone for $m=0.35$
corresponding to the hidden topological phase in the 3D bulk limit.
Large intensity of $F^{\text{Total}}_{\textbf{k}}$ occurs along
the loop corresponding to the overlapping region
of two in-gap states localized on the top and bottom surfaces, respectively.
({\bf b}) Band structure along the high symmetry directions.
The highest energy occupied (HO) band
is indicated by a blue line.
({\bf c})-({\bf e}) Distribution of the Berry curvature near the loop with high intensity.
Here $F^{\text{HO}}_{\textbf{k}}$ indicates the Berry curvature from the HO band and
$F^{\text{Diff}}_{\textbf{k}}\equiv F^{\text{Total}}_{\textbf{k}}-F^{\text{HO}}_{\textbf{k}}$.
({\bf f}) Layer-resolved wave function amplitudes for the HO band
along the direction marked by the orange arrow in ({\bf a}).
Here $|\Psi^{\text{Top}}_{\text{HO}}|$
($|\Psi^{\text{Bottom}}_{\text{HO}}|$) indicates the wave function amplitude
on the top (bottom) layer and
$|\Psi^{\text{Middle}}_{\text{HO}}|^{2}\equiv1-|\Psi^{\text{Top}}_{\text{HO}}|^{2}-|\Psi^{\text{Bottom}}_{\text{HO}}|^{2}$.
} \label{fig:surfaceQPT}
\end{figure*}

\begin{figure*}[t]
\centering
\includegraphics[width=16 cm]{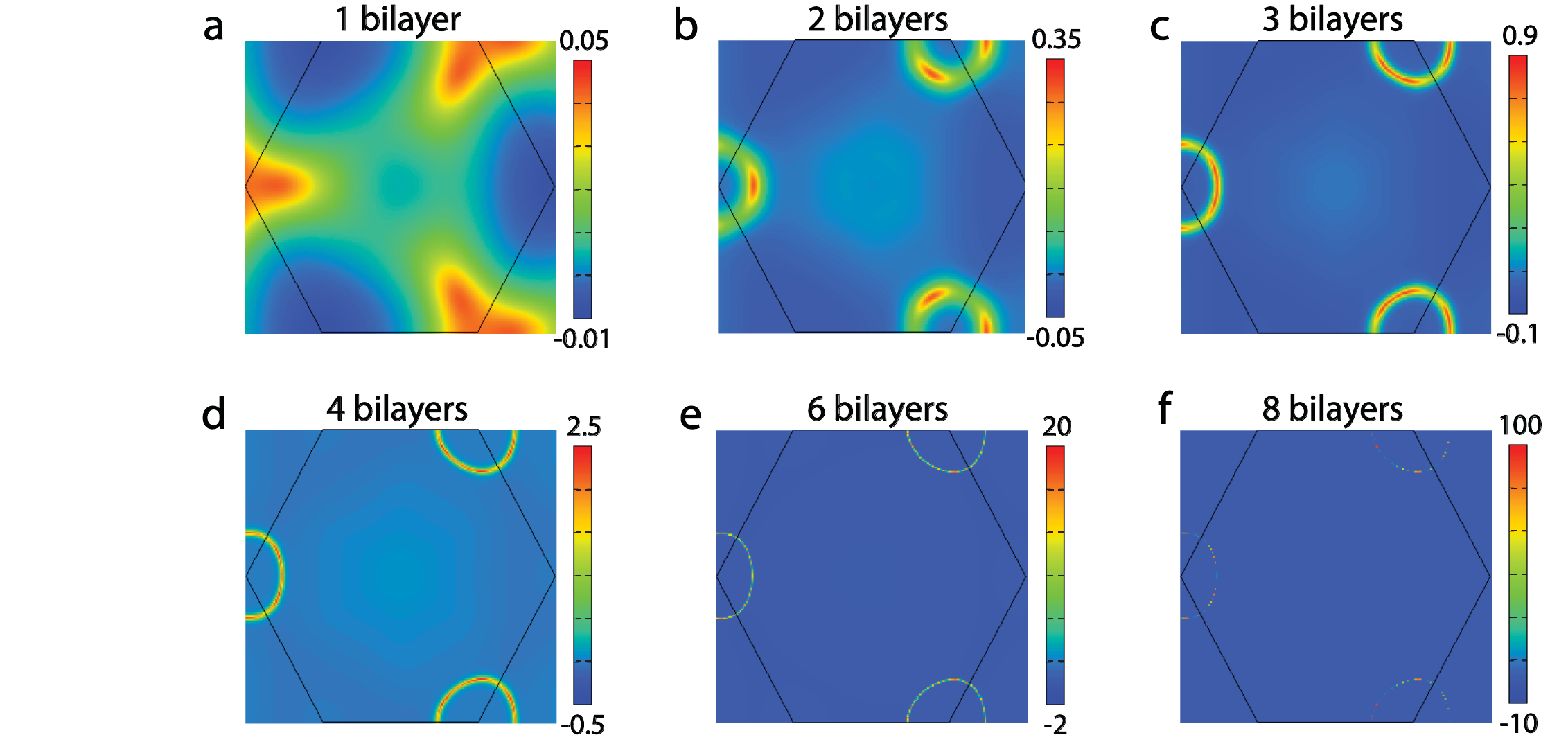}
\caption{
{\bf Layer thickness dependence of the distribution of Berry curvature.}
Distribution of Berry curvature of occupied bands in the momentum space for
({\bf a}) 1 bilayer. ({\bf b}) 2 bilayers.
({\bf c}) 3 bilayers. ({\bf d}) 4 bilayers.
({\bf e}) 6 bilayers. ({\bf f}) 8 bilayers.
Here we choose $m=0.35$ corresponding to the hidden topological phase
in the bulk limit.
To compute the Berry curvature, we assumed the half-filled condition
locally in each momentum $\textbf{k}$.
As the number of bilayer increases, the Berry curvature
is peaked in a narrower region. This reflects
the fact that the enhanced Berry curvature occurs
due to the overlap between two surface states, which are
exponentially localized on the top and bottom surfaces, respectively.
} \label{fig:Berrycurvature}
\end{figure*}

\begin{figure*}[h]
\centering
\includegraphics[width=11 cm]{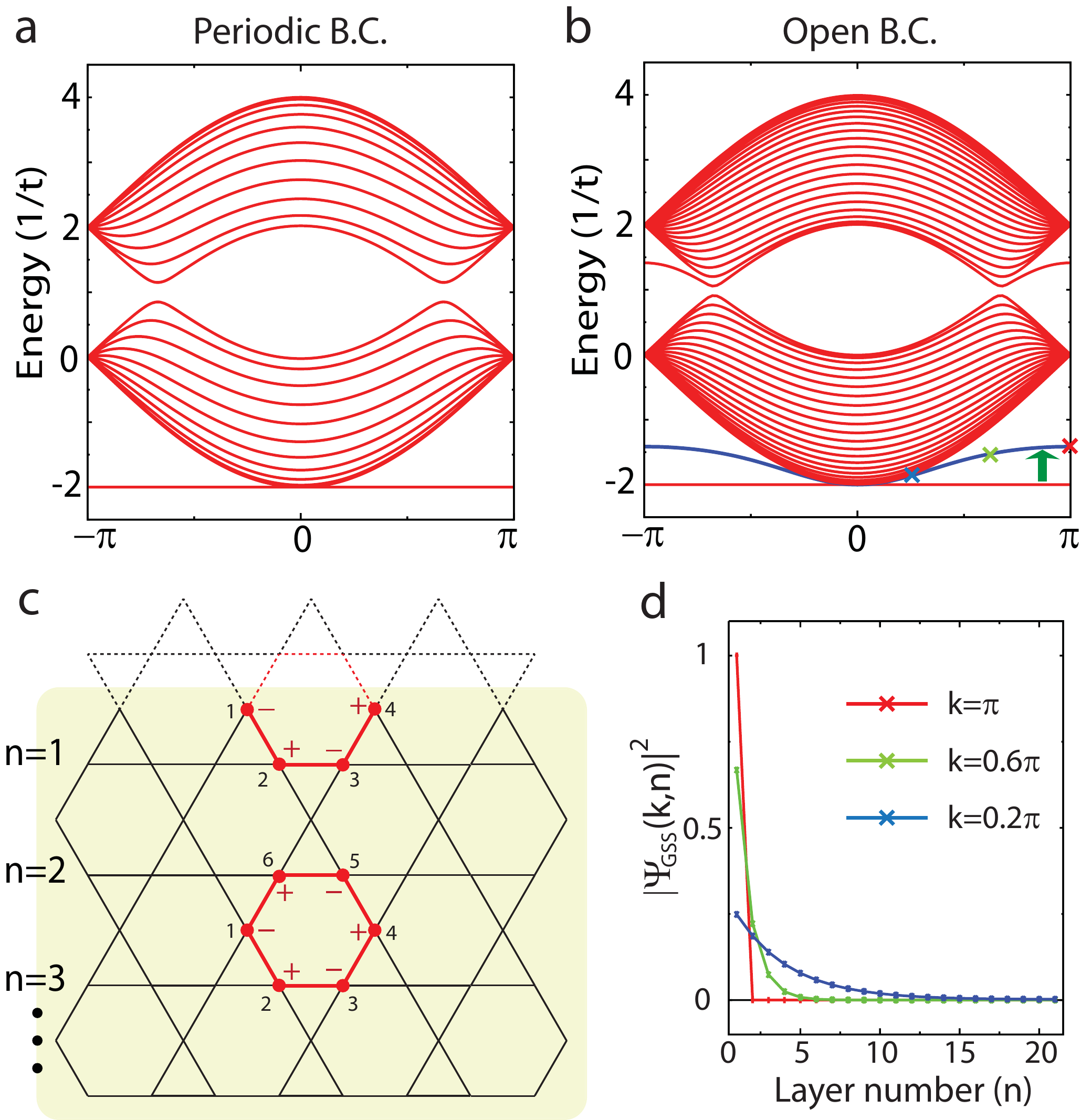}
\caption{
{\bf Localized edge states due to geometrical structure of kagome lattice.}
Energy spectrum of a kagome ribbon under the periodic
BC ({\bf a}) and the open BC ({\bf b}).
({\bf c}) The kagome ribbon has a sawtooth shape (a straight line shape) on the top (bottom) edge
while translational invariance is maintained along the horizontal direction.
The kagome ribbon can be obtained by vertical stacking of 1D layers (labeled by $n$) composed
of horizontally connected up-triangles.
Under the open BC, we can see an emerging localized
state marked by a blue line in ({\bf b}), i.e.,
the geometry-induced surface state (GSS), which was a part
of flat bands under the periodic BC.
Due to the geometrical structure of the kagome lattice,
each hexagon supports a localized state which gives rise to
flat bands in the momentum space.
An example of such a localized state confined around a hexagon is indicated by
a red closed loop in ({\bf c}).
({\bf d}) Spatial distribution of the squared wave function of the GSS ($|\Psi_{\text{GSS}}|^{2}$)
along the vertical direction.
The GSS is exponentially localized on the top edge with sawtooth shape.
}
\end{figure*}

\begin{figure*}[h]
\centering
\includegraphics[width=15 cm]{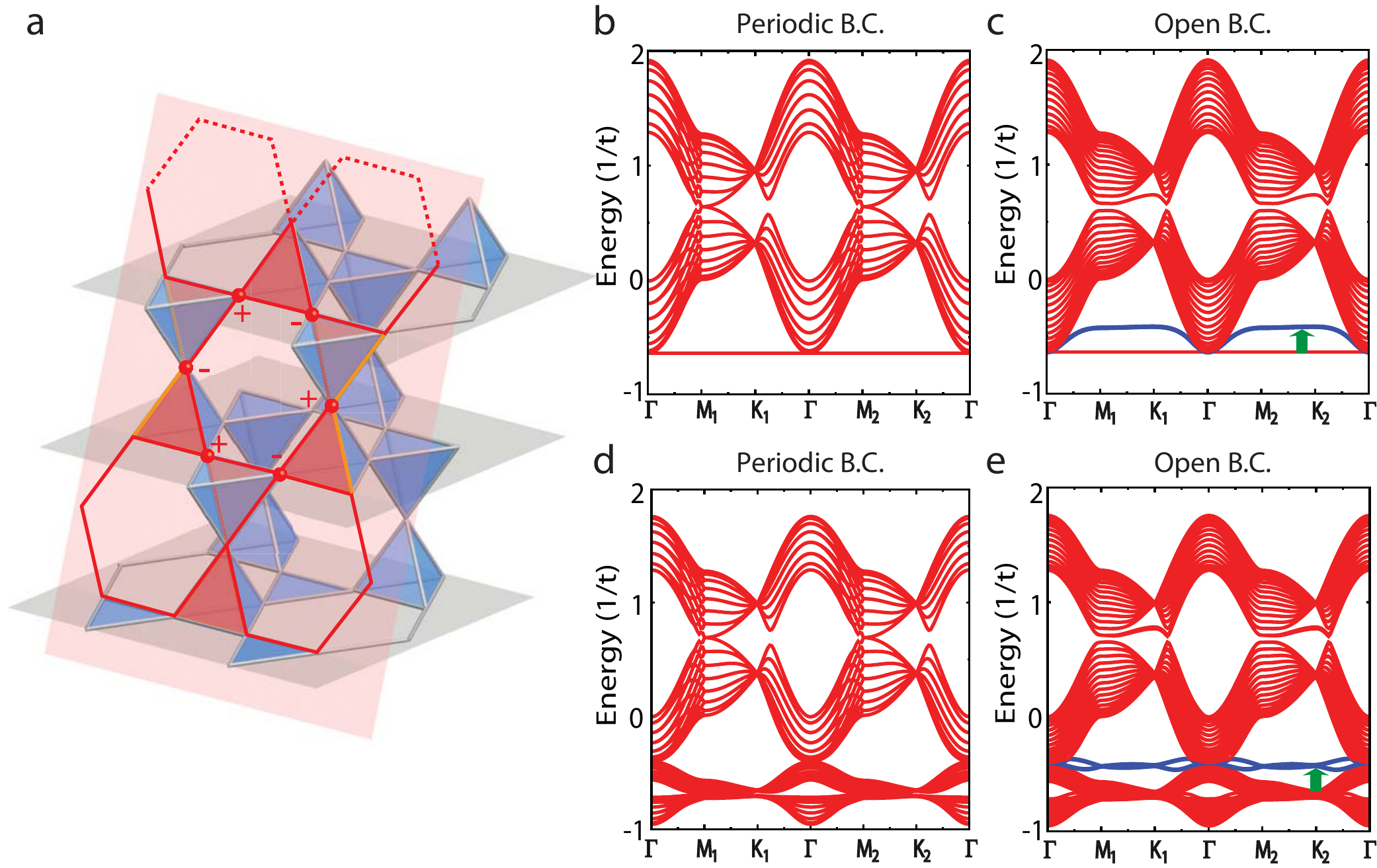}
\caption{
{\bf Localized edge states due to geometrical structure of pyrochlore lattice.}
({\bf a}) Localized states in a [111] film terminated by a triangular (kagome)
lattice on the top (bottom) surface.
On the top surface, there
are broken hexagons, which generate GSS which are exponentially localized
near the top surface layer.
({\bf b, c}) Band structure of nearest neighbor hopping Hamiltonian
for a [111] film under the periodic BC and under the open BC.
({\bf d, e}) Inclusion of the second nearest neighbor hopping
and various effective spin dependent
hopping processes induces the dispersion of low energy bands which are
completely flat when only the nearest neighbor hopping is considered.
The GSS still appears as long as it exists within a gap.
}
\end{figure*}


\


\


\end{document}